\documentstyle[preprint,revtex]{aps}
\def\v{\vec{v}}
\def\w{\vec{\omega}}
\def\rot{{\rm rot}}
\def\div{{\rm div}}
\def\r{{\bf r}}

\def\hi{\hat{\bf i}}
\def\hz{\hat{\bf z}}
\def\tvi{\tilde v_{i}}

\def\wj{\omega_{j}}

\def\wx{\omega_{x}}

\tightenlines
\begin{document}
\draft
\preprint{titcmt-1992-11}
\begin{title}
Non-Gaussian distribution
in Random advection dynamics
\end{title}
\author{Y-h. Taguchi}
\begin{instit}
 Institut f\"ur Festk\"orperforschung,
 Forschungszentrum J\"ulich,\\
 D-5170 J\"ulich, Germany\\
 and\\
 Department of Physics,
Tokyo Institute of Technology,\\
Oh-okayama, Meguro-ku,
 Tokyo 152, Japan\cite{present}
\end{instit}
\author{Hideki Takayasu}
\begin{instit}
 Department of Earth Science, Kobe University, Kobe 657, Japan.
\end{instit}
\receipt{}
\begin{abstract}

Simulations of vortex tube dynamics reveal that the
non-Gaussian nature of turbulent fluctuation
originates in the
effect of random advection.
A similar non-Gaussian distribution is found numerically
in a simplified statistical model of random
advection.
An analytical solution is obtained in the mean-field case.
\end{abstract}
\narrowtext

In the study of statistical physics for non-equilibrium
systems the deviation from the
Gaussian distribution has been the
central issue.
We are expecting the existence of universal mechanisms of producing
non-Gaussian distributions as in the case of thermal equilibrium systems,
however, our knowledge has been still in the elementary level.

Non-Gaussian probability distribution function (NGPDF) is
especially important in fluid turbulence.
Turbulence is a typical far-from-equilibrium phenomena
as it is characterized by the energy cascade
which violates both the detailed balance and equi-partition.
The appearance of NGPDF is so common and we can not construct any theory
without taking the non-Gaussian nature
into account\cite{Batchelor,Monin}.
Latest technique of direct observation of turbulence\cite{Goldburg} and
also direct numerical integration of Navier-Stokes equation\cite{Kida} are
clarifying the details NGPDF from experimental viewpoints.

Recently, theoretical investigation to NGPDF in turbulence
is attracting much attention. An approach is to assume the
multifractality in the geometry of turbulent velocity field as an
ideal limit\cite{Benzi},
and another one called the mapping closure conjectures the
existence of a kind of smooth map which describes the
time evolution of statistical quantities\cite{She91,Kraichnan}.

Yakhot et al.\cite{Yakhot} also present a phenomenological
excellent approach
in order to derive a NGPDF for vorticities in the
fluid turbulence. Although it is fairly reasonable,
physical meanings in their assumptions are not clear.

 In this paper we first introduce a numerical model of
vortex dynamics on lattice and show that the vorticity distribution in a
turbulent steady state is actually far from
Gaussian but closer to a symmetric exponential distribution.
In order to clarify the origin of
this NGPDF we modify the dynamics so that vortex
tubes move randomly by passive advection.
It is shown that this modification does not seem to affect
the vorticity distribution,
which implies that the random advection is playing an important role.
Then, we focus our attention to the effect of random advection
and introduce a
random scalar advection model on lattice.
Surprisingly the scalar also follows a distribution which
is almost identical to the
former NGPDFs.
A mean-field version of the scalar model is then solved
analytically showing a clear evidence of NGPDF.

Let us first introduce our lattice vortex model\cite{Tag}
which is equivalent to the following set of vorticity equations of
incompressible fluid in the continuum limit
\begin{mathletters}
\begin{eqnarray}
\frac{\partial \w}{\partial t} + (\v \cdot \nabla ) \w &=& (\w \cdot \nabla)\v
+ \nu \Delta \w, \label{vor}\\
\rot \v &=& \w, \label{rot}\\
\div \v &=& 0,  \label{div}\\ \nonumber
\end{eqnarray}
\end{mathletters}

where $\v, \w$ and $\nu$ are the velocity, vorticity and viscosity,
respectively.
We assign vorticities on a simple cubic lattice.
Every bond has only one vorticity component along its direction.
This means that  an $x$-bond has only $x$ component of vorticity,
i.e., a bond can be viewed as a vortex tube.
For a given configuration of vorticities the vorticity field is calculated by
using the Biot-Savier law (equivalent to
eqs.(\ref{rot}) and (\ref{div}).
The fluid velocity of a bond at $\r,\v(\r)$, is estimated at the mid-point of
the bond.
Dynamics is defined so as to satisfy
the vorticity equation(\ref{vor}).

Due to the advection term $(\v \cdot \nabla) \w$ in eq.~(\ref{vor})
the vorticity flows with the fluid as shown
by the name of Kelvin's theorem\cite{Landau},
the vorticity $\omega_z$ at a $z$-bond is transported to its six
neighboring $z$-bonds.
By this effect the value of $\omega_z$ at $\r \pm \hi$
is increased by $\pm J_{iz} \Delta t / 2$,
where $\hi$ is a unit vector directing either $x$, $y$ or $z$ direction,
and $J_{ij}$ is the flux of vorticity, $\tvi \wj$.
The coefficient $1/2$ of the vorticity flux and the signs $\pm$ are
introduced to keep the spatial symmetry.
The first term on the right hand side of eq.(\ref{vor})
shows the vortex stretching term,
so we have to modify the vorticities around the
advectively transported bonds so that modified vorticities
make loops as shown in Fig.\ref{dyna}.
In this procedure we add $\mp  J_{iz} \Delta t/2$
to the bonds at $\r \pm \hi/2 \pm \hz/2$,
namely, $x$ and $y$ component appear due to the
elongation of
vortex tube.
The diffusion term in eq.~(\ref{vor}) is
discretized by the usual finite difference method.
It has been shown that this
dynamics is
equivalent to eq.(\ref{vor}) in the continuum limit and
numerical simulation of this vortex tube dynamics on a periodic
$24 \times 24 \times 24$ lattice has reproduced many basic
properties of fluid turbulence\cite{Tag}.

As for the probability distributions we obtained the following results:

1.The velocity components (for example $v_x$) follows
nearly Gaussian.

2. The distribution of differentiated quantities, such as
$\partial v_x /\partial x, \partial v_x / \partial y $
and $\omega_x$, are non-Gaussian and close
to exponential (see Fig.\ref{wx}).

3. By removing lower wave number components, the
distribution of $v_x$ also becomes closer to
an exponential as firstly found by She et al\cite{She88}.

We also observe a kind of
conditional distributions for
velocities on bonds whose vorticities are
in a fixed range.
The distribution are nearly identical both for
large $\omega$ and for small $\omega$,
indicating that the velocity on a bond
is nearly independent of its vorticity strength.

We now modify our model to seek the origin of the NGPDF.
The modification is to randomize the velocity field at each time step
keeping its distribution.
Namely, we do not use Biot-Savier law, but
the vorticities are transported and elongated by random
advection.
The probability distribution of $\omega$ after several hundreds time steps is
quite
similar to that of original lattice vortex model (see Fig.\ref{wx}).
This result clearly shows that the appearance of NGPDF
is independent of the details of velocity field.
As suggested by Sinai and Yakhot\cite{Sinai,Yakhot} NGPDF may be
caused by random advection,
and if so, we can expect
a large universality class including another exponential-like distribution in
thermal convection flow\cite{Siggia,Yanagita}.

In order to see the effect of random advection more clearly
we introduce a scalar model on lattice.
The model is governed by the
following stochastic equation
\begin{equation}
\omega (r_0,t + \Delta t) = \omega (r_0,t) - v(r_0,t) \omega (r_0,t) \Delta t
+ \sum_{r} P(r_0,r,t) v(r,t) \omega (r,t) \Delta t,
\label{eq:adv}
\end{equation}

where $v(r,t)$ and $P(r_0,r,t)$ are independent random variables.
$v(r,t)$ takes non-negative values, and $P(r_0,r,t)=1$
when a flow from site $r$ to $r_0$ exists at time $t$,
otherwise $P(r_0,r,t)=0$ and it is normalized as
$\sum_r P(r_0,r,t)=1$. The one-dimensional version of eq.(\ref{eq:adv})
is defined by the special case that either $P(r_0,r_0-1,t)$ or
$P(r_0,r_0+1,t)$ is equal to 1 with probability $1/2$.
We perform the simulation on a 1-dimensional
lattice of size $10000$ with the periodic boundary condition.
The random number $v(r,t)$ is in the range of $[0,0.5]$ and
$\Delta t =0.5$.
For a given random initial condition of ${\omega (r,0)}$
we observe the distribution of $\omega(r,t)$ at
sufficiently large $t$.
The result is also plotted in Fig.\ref{wx}.
The distribution is far from Gaussian and is very close to those of
former cases.
Now it seems obvious that the origin of NGPDF is in the
simple random advection transport.

A mean-field version of eq.(\ref{eq:adv}) can be solved analytically.
We consider the situation that a site is interacting
with a mean-field site.
 To avoid unimportant complexity we assume the case
that the random number $v$ takes either 0 or $j_0/\Delta t$
with probability $1/2$ ($j_0 \in [0,1]$).
In this situation the stochastic eq.(\ref{eq:adv}) becomes
\begin{equation}
\omega (t+t_0) = \left \{ \begin{array}{ll}
 \omega (t), & {\rm Prob.} 1/4 \\
(1-j_0) \omega (t), & {\rm Prob.} 1/4 \\
\omega (t) + j_0 \omega_M, & {\rm Prob.} 1/4 \\
(1-j_0) \omega (t) + j_0 \omega_M, & {\rm Prob.} 1/4
\end{array}, \right.
\label{eq:mean}
\end{equation}
where $\omega_M$ is an independent random number having the same
distribution as $\omega (t)$.
By introducing the characteristic function
$$
Y(\rho,t)=\int^{\infty}_{-\infty} e^{i\rho \omega} P(\omega,t) d \omega,
$$
eq.(\ref{eq:mean}) becomes
\begin{equation}
Y(\rho, t+\Delta t) = \frac{1}{4} \{ Y(\rho,t) + Y(\rho-\rho j_0,t)\}
\{ 1 + Y ( \rho j_0,t) \}.
\label{eq:mean2}
\end{equation}

Taylor expansion of eq.(\ref{eq:mean2})
in terms of $\rho$ gives a set of equations
for the moment functions $\{ < \omega^n > \}$,
and it can be easily shown that a steady state exists when $< \omega > \neq 0$.

The steady state solution is obtained with the aid of algebraic
calculation by computer.
In Fig.\ref{fig:j0} we plot the cumulants for
orders up to $7$ together with those for one-sided exponential
distribution for comparison.
Higher order cumulants are not zero in any case and
showing a tendency to diverge as the order goes to infinity,
which clearly demonstrates that the steady state distributions are
not Gaussian.

In the special case of $j_0 = 1/2$ we can estimate the
functional form of $Y(\rho)$ as follows. By denoting
$Y_n = Y(\rho 2^n)$ eq.(\ref{eq:mean2}) becomes
\begin{equation}
Y_{n+1}-Y_n = 2 \cdot \frac{Y_n^2 -Y_n}{3 - Y_n}.
\label{eq:j2}
\end{equation}
Supposing that $n$ can be a continuous number and we
approximate the left hand side by a derivative $dY/dn$.
Then eq.(\ref{eq:j2}) can easily be
integrates and we have the following cubic equation for $Y(\rho)$ with
$< \omega > = 1$,
\begin{equation}
\rho^2 Y(\rho)^3 + (Y(\rho)-1)^2 =0,
\end{equation}
which can be solved explicitly by Cardano's formula.
The asymptotic behavior for $\rho \rightarrow \infty$ is given as
\begin{equation}
Y(\rho) \propto \rho^{-2/3}.
\end{equation}

This power law decay is very different from the case
of a Gaussian
distribution where the characteristic function decays faster than any power
of $\rho$. In the case of exponential distribution the characteristic
function is given by $1/(1 + i \rho)$ (asymmetric case)
or $1/(1+\rho^2)$ (symmetric case), which also shows
a power law decay as $\rho \rightarrow \infty$.
In this sense our distribution is far from Gaussian,
but has a similarity to
the exponential distribution.

For smaller $j_0$ the cumulants are smaller, and
in the limit of $j_0 \rightarrow 0$
it can  be shown that the normalized cumulant of order
$n$, $< \omega^n >_c / < \omega^2 >_c^{n/2}$,
vanishes for $n \geq 3$.
Namely, the
distribution converges to a Gaussian in this limit,
which agrees with the well-known fact that
the distribution function
satisfying the usual diffusion equation in the
continuum limit is a Gaussian since eq.(\ref{eq:adv})
becomes a usual diffusion equation in the limit
of $\Delta t \rightarrow 0$.
This indicates that the finiteness of
$\Delta t $ is very
important for the appearance of NGPDF in our random advection
dynamics.

One may think that
these results for $\Delta t \neq 0$ are meaningless
because we must take the limit of $\Delta t \rightarrow 0$ in
order to recover eq.(\ref{vor}).
This is a wrong argument neglecting the finiteness of correlation time in
the real dynamics with $\Delta t =0$.
It is known that the fastest mode in real turbulent
flows has a characteristic time approximately given by
$\sqrt{\nu / \varepsilon}$,
where $\varepsilon$ is the mean energy dissipation rate
\cite{visco}.
For time scales less than this value advections can not be treated
as random.
In our random advection models $\Delta t$ should be viewed as the
correlation time in real dynamics, therefore, it should take a finite value.

Our theoretical approach is successful only in the case
of $< \omega > \neq 0$,
while observed distributions are symmetric with $< \omega > = 0$.
At present we have no rigorous way of
connecting these two cases,
but we are now considering the effect of spatial fluctuation as
the key of solving this difficulty.
Even in the case of $< \omega > =0$ averaged over
the whole space, $<\omega >$ can be nonzero if the average is taken over
a finite area due to the spatial fluctuation.
So, if the distribution of $\omega$ is determined rather
locally not using the information from the whole space,
then the distribution for $< \omega > \neq 0$ may have
direct consequence to
the real distributions.

In summary, we showed that the random
advection dynamics creates far-from Gaussian
fluctuations whose distribution functions
are
closer to the exponential.
As shown in the discussions of the statistical
scalar model on lattice and its mean-field analysis,
the mechanism of NGPDF is so simple that a wide application
may be expected.

H.T. thanks M. Takayasu for useful discussions.
NEC software (c) is also acknowledged for allowing Y.T. to use
EWS-4800/220, with which all calculations were performed.
 This work is supported by Grant-Aid for Scientific
Research on Priority Areas,
"Computational Physics as a New Frontier in Condensed Matter Research",
from the Ministry of Education, Science and Culture, Japan.

\figure{Dynamics of vortex tubes. Direction of arrows
indicates the sign. (See text)\label {dyna}}
\figure{Probability distributions
 of $\wx$; $\Diamond$  for true vortex dynamics
\protect\cite{Tag}, $+$ for random advection model of
vortex tubes, $\Box$ for the 1-dimensional scalar model.
They are averaged over ten samples.
Error bars stand for standard deviations.
$\wx$ is normalized so as to have variance of unity.
\label{wx}}
\figure{Higher order cumulants $\langle \omega^n \rangle_c$.
$+$ for the mean-field model;
$j_0=0.1,0.7$ and $0.9$ from bottom to top.
The exponential distribution
is  shown for comparison($\Diamond$).
\label{fig:j0}}

\begin{references}
\bibitem{visco} According to the Kolmogorov's theory\protect\cite{Batchelor},
the smallest scale which appears in the homogeneous
isotropic turbulence is estimated by $( \nu^3 / \varepsilon )^{1/4}$.
The characteristic time scale corresponding to this scale is
$\sqrt{\nu/\varepsilon}$.
\end{references}
\end{document}